\documentclass[aps,twocolumn,pra,superscriptaddress,nofootinbib,amsmath,amssymb,floats,floatfix,english]{revtex4-1}
\usepackage{polski}
\usepackage[utf8]{inputenc}
\usepackage[T1]{fontenc}
\usepackage{amsmath}
\usepackage{amssymb}
\usepackage{ifthen}
\usepackage{graphicx}
\usepackage{times}
\usepackage{babel}
\usepackage{color}
\usepackage{bm}
\usepackage{pstricks}

\usepackage[percent]{overpic}
\usepackage{tikz}
\usepackage{ulem}

\DeclareMathSizes{7}{7}{5}{4} 
\DeclareMathSizes{6}{6}{3}{2} 
\DeclareMathSizes{5}{5}{3}{2}

\DeclareMathSizes{5}{5}{2}{2}

\newcommand{\op}[1]{\widehat{#1}}
\newcommand{\dagop}[1]{\widehat{#1}^{\dagger}}
\newcommand{\bo}[1]{{\mathbf{#1}}}
\newcommand{\mc}[1]{{\mathcal{#1}}}
\newcommand{\wt}[1]{{\widetilde{#1}}}
\newcommand{\wb}[1]{{\overline{#1}}}

\newcommand{\nonu}{\nonumber}

\newlength{\templength}

\newenvironment{eq}{\begin{equation}}{\end{equation}}

\newcommand{\eqn}[1]{(\ref{#1})}

\renewcommand{\eq}[2]{\begin{equation}\label{#1}#2\end{equation}}

\newcommand{\eqa}[2]{\begin{eqnarray}\label{#1}#2\end{eqnarray}}

\usepackage{lmodern}

\newcommand{\ve}{\varepsilon}

\begin{document}
\title{Classical fields in the one-dimensional Bose gas: \\
 applicability and  determination  of the optimal cutoff   
}

\author{J. Pietraszewicz}
\affiliation{Institute of Physics, Polish Academy of Sciences, Aleja Lotnik\'ow
32/46, 02-668 Warsaw, Poland}
\email{pietras@ifpan.edu.pl, deuar@ifpan.edu.pl}

\author{P. Deuar}
\affiliation{Institute of Physics, Polish Academy of Sciences, Aleja Lotnik\'ow
32/46, 02-668 Warsaw, Poland}

\date{\today}

\begin{abstract}

To finalize information about
the accuracy of the classical field approach 
for the 1d Bose gas, the lowest temperature quasicondensate was studied by  comparing the extended Bogoliubov model of Mora and Castin, to its classical field analogue. 
The parameters for which the physics is well described by matter waves are now presented for all 1d regimes, and
concurrently, the optimal cutoff that best matches all observables together is also provided. 
This cutoff rises strongly with density when the chemical potential is higher than the thermal energy to account for kinetic energy.
As a consequence, clouds that reach this coldest quantum fluctuating regime are better described using a trap basis than plane waves. This contrasts with higher temperature clouds for which the basis choice is less important.
In passing, estimates for chemical potential, density fluctuations, kinetic and interaction energy in the low temperature quasicondensate are obtained up to several leading terms.
\end{abstract} 

\maketitle

%%%%%%%%%%%%%%%%%%%%%%%%%%%%%%%%%%%%%%%%%%%%%%%%%%%%%%%%%%%%%%%%%%%%%%%%%%%%%%%%%%%%%%%%%%%%%%%%%%%%%%%%%%%%%%%%%%%%%%%%%%%%%%%%%%%%%%%%%%%%%%%%%%%%%%%%%%%%%%%%%%%%%%%%%%%%%%%%%%%%%%%%%%%%%%%%%%%%%%%%%%%%%%%%%%%%%%%%%%%%%%%%%%%%%%%%%%%%
\section{Introduction}
\label{INTRO} 

The question of under what conditions the classical field description of ultracold gases is accurate has been widely discussed in the field \cite{FINESS-Book-Davis,FINESS-Book-Cockburn,FINESS-Book-Brewczyk,FINESS-Book-Wright,FINESS-Book-Ruostekoski,Brewczyk07,Blakie07,Blakie08,Bradley05,Davis06,Cockburn11a,Pietraszewicz15,Pietraszewicz18a}. 
 In parallel, the companion question is what high energy cutoff should be chosen for best results \cite{Witkowska09,Brewczyk04,Zawitkowski04,Sinatra02,Rooney10,Bradley05,Blakie07,Sato12,Cockburn12,Pietraszewicz15,Pietraszewicz18a}. 
The importance of these matters stems from the widespread utility of the method for nonperturbative and thermal calculations \cite{FINESS-Book-Davis,Blakie08,Kagan97,Sadler06,Weiler08,Martin10a,Witkowska11,Cockburn11a,Sabbatini12,Karpiuk12,Gring12,Parker13,Liu16,Tsatsos16,Liu18} and its interpretation in terms of matter waves.

Previous work in the 1d Bose gas gave detailed quantitative answers to these questions for most degenerate temperatures \cite{Pietraszewicz15,Pietraszewicz18a}, by comparing classical field ensembles with the exact Yang-Yang results \cite{Yang69,Kheruntsyan05,PDJPYang17}.
However, the full picture was not obtained because there were technical difficulties in assessing the colder quasicondensate for which quantum fluctuations become significant. In practice this meant that 
when thermal energy $k_B T$ is comparable to or lower than the chemical potential, the status of the classical field description of the centre of a gas cloud was unclear.

Here, to obtain complete coverage,
we re-analyze the case of the quasicondensate by comparing the extended Bogoliubov theory of Mora and Castin \cite{Mora03} with its classical field counterpart \cite{Sinatra12}. 
The present analysis conforms with the previous results but also extends the determinations down to zero temperature. In this way,  
a comprehensive assessment across 
all regimes of the 1d Bose gas is now provided in this work. 

 The structure of the paper is as follows:
 Sec.~\ref{assum} gives the background information,  while
 Sec.~\ref{BOG} describes the extended Bogoliubov description and its classical field version that we will use to study the quasicondensate.
 Sec.~\ref{FOM-OPT} explains how classical field accuracy will be judged there.
 Sec.~\ref{RMS} compares full quantum and classical field predictions and gives the main results, i.e. the limits of the matter wave region and the optimal cutoff prescription.
 Sec.~\ref{ruleTh} discusses the physical reasons for the high cutoff found in the quantum fluctuating region, and its consequences. 
 We summarize in Sec.~\ref{CONCLUSIONS}. Additional technical details are given in the appendices, as well as a number of analytic estimates for the main observables in the Bogoliubov regime.

\section{Background}
%%%%%%%%%%%%%%%%%%%%%%%%%%%%%%%%%%%%%%%%%%%%%%%%%%%%%%%%%%%%%%%%%%%%%%%%%%%%%%%%%%%%%%%%%%%%%%%%%%%%%%%%%%%%%%%%%%%%%%%%%%%%%%%%%%%%%%%%%%%%%%%%%%%%%%%%%%%%%%%%%%%%%%%%%%%%%%%%%%%%%%%%%%%%%%%%%%%%%%%%%%%%%%%%%%%%%%%%%%%%%%%%%%%%%%%%%%%%%
\label{assum}

 The interest in a precise characterization of the classical field description is twofold. 

 The \textsl{practical aspect} is the usage of the classical field method to simulate dynamics. Many
 kinds of non-perturbative phenomena have become accessible experimentally in recent years \cite{Donadello14,Serafini15,Navon15,Liu16,Tsatsos16}, but
 in a vast range of cases, only classical fields remain
 tractable for very large systems. Furthermore, they also give access to predictions
 for single experimental runs \cite{Kagan97,Duine01,Lewis-Swan16,Blakie08,FINESS-Book-Cockburn,FINESS-Book-Davis,Javanainen13,Lee14,FINESS-Book-Wright}.
 Several flavors of c-fields have been developed \cite{Brewczyk07,Blakie08,Proukakis08,Gardiner03,Sinatra02} and applied to 
defect seeding and formation \cite{Lobo04,Weiler08,Bradley08,Bisset09a,Damski10,Witkowska11,Karpiuk12,Simula14,Liu16},
quantum turbulence \cite{Berloff02,Parker05,Wright08,Tsatsos16},
the Kibble-Zurek mechanism \cite{Sabbatini12,Swislocki13,Witkowska13,Anquez16},
nonthermal fixed points \cite{Nowak12,Schmidt12,Karl13,Nowak14}, vortex dynamics \cite{Bisset09a,Karpiuk09,Rooney10},
the BKT transition \cite{Bisset09a},
evaporative cooling \cite{Marshall99,Proukakis06,Witkowska11,Liu18}, and more.

 All c-field varieties tend to suffer from ambiguity regarding the best choice of high-energy cutoff,
 because predictions of observables can depend sizeably on the cutoff choice.
 A range of prescriptions for choosing cutoff
 have been developed \cite{Brewczyk04,Zawitkowski04,Karpiuk10,Blakie08,Rooney10,Bradley05,Witkowska09,Sinatra12,Cockburn12,Pietraszewicz15,Pietraszewicz18a}
 but generally no particular choice is ideal. For example, a cutoff that leads to correct predictions of density
 and one additional observable will inaccurately describe other quantities \cite{Pietraszewicz18a}.

The \textsl{physical aspect} of characterizing ``classical'' field descriptions is that they
describe the physics of matter waves, while neglecting effects due to particle discretization. In the ultracold atom domain this is not at all the same as so-called \emph{classical physics}. However, it does mean that wave-particle duality is insignificant whenever a classical field description is good. Hence, by studying its accuracy, one can show the regimes in which wave-particle duality is relevant or not.

Quantitative studies were begun in \cite{Pietraszewicz15,Pietraszewicz18a} and are
continued here.
A figure of merit min$RMS$, first identified in \cite{Pietraszewicz18a}, 
bounds the discrepancy in all the standard observables, 
 and provides a cutoff value opt$f_c$ that minimizes inaccuracies.  A good classical wave description will only be present if there is some cutoff choice  that leads to small discrepancy  in \emph{all} the relevant observables simultaneously. Below 10\% is a reasonable value, since experimental precision is also of this order.

As in the analyses of \cite{Pietraszewicz15,Pietraszewicz18a}, we will consider locally uniform sections of the 1d gas in the grand canonical ensemble. The latter 
corresponds to thermal and diffusive contact between neighboring sections. Such assumptions enable wider usage of the results for non-uniform gases via a local density approach. 
We also assume that the gas sections are large enough to be in the thermodynamic limit with regard to the observables that will be considered. In the Bogoliubov region, the limiting ones are kinetic energy $\ve$ and the phase coherence.

Such a uniform 1d Bose gas section is fully characterized by only two parameters:
The dimensionless interaction strength $\gamma$ and dimensionless temperature $\tau_d$:
\eq{para}{
\gamma=\frac{m g}{\hbar^2 n}; \qquad
\tau_d=\frac{T}{T_{d}} = \frac{1}{2 \pi} \frac{m k_B }{\hbar^2 } \frac{ T}{n^2}.
}
Here $n$ is the density,  $T$  the temperature, $m$ the particle mass, and $g$ the contact interaction strength. 
When temperature reaches the quantum degeneracy temperature $T_d$, 
there is about one particle per thermal de Broglie wavelength $\Lambda_T=\sqrt{\frac{2 \pi \hbar^2}{m k_BT}}$.

The physical regime that is particularly relevant for the analysis here is the \textsl{quasicondensate} lying in the range $\tau_d^2\lesssim\gamma\lesssim1$ in which density fluctuations are small, and the relation
\eq{muT}{
\mu \approx k_BT\ \frac{\gamma}{2\pi\tau_d}
}
holds. 
The quasicondensate consists of two physically different regions characterized by the dominance of either:
\begin{itemize}
\item[-] \textsl{Thermal fluctuations} when $k_BT \gtrsim \mu$, or
\item[-] \textsl{Quantum fluctuations} when $k_BT \lesssim \mu$.
\end{itemize}
This distinction makes a large difference for classical field accuracy and cutoff dependence. 

Low temperatures with $k_BT\lesssim \mu$ were difficult to access 
using the methods employed previously \cite{Pietraszewicz18a}.
Convergence to the equilibrium state in the thermodynamic limit became very slow there, both for the iterative algorithm that is used to solve the Yang-Yang integral equations, and also for the generation of classical field ensembles via Metropolis \cite{Witkowska10}  or SPGPE \cite{Gardiner03}. This slowness was compounded by the growth of the numerical lattices as temperature falls.

%%%%%%%%%%%%%%%%%%%%%%%%%%%%%%%%%%%%%%%%%%%%%%%%%%%%%%%%%%%%%%%%%%%%%%%%%%%%%%%%%%%%%%%%%%%%%%%%%%%%%%%%%%%%%%%%%%%%%%%%%%%%%%%%%%%%%%%%%%%%%%%%%%%%%%%%%%%%%%%%%%%%%%%%%%%%%%%%%%%%%%%%%%%%%%%%%%%%%%%%%%%%%%%%%%%%%%%%%%%%%%%%%%%%%%%%%%%%%
\section{Quasicondensate description}
\label{BOG}

The quasicondensate is very well described by the extended Bogoliubov model given by Mora and Castin \cite{Mora03}, across a wide range of temperatures. 
The model does not assume a single phase-coherent dominant condensate mode like standard Bogoliubov \cite{Bogo47, Bogo58}, but makes an expansion in small density fluctuations instead. 
Section~\ref{BOGQ} summarizes the resulting fully quantum description for the uniform gas which will be our baseline for comparison, while Sec.~\ref{BOGC} describes the corresponding classical field description.

\subsection{Extended Bogoliubov model for a uniform 1d gas}
\label{BOGQ}

The boson field $\op{\Psi}(x)$ in this model is expressed as
\eq{psiBog}{
\op{\Psi}(x) = e^{i\op{\theta}(x)}\sqrt{\op{\rho}(x)},\quad
\op{\rho}(x) = \rho_0 + \delta\op{\rho}(x),
}
with the help of operators for the phase $\op{\theta}(x)$ and density fluctuations $\delta\op{\rho}(x)$. 
The quantity $\rho_0$ is the lowest order density estimate obtained from the Gross-Pitaevskii solution.  
The gas section of length $L$ is discretized into sites of length $\Delta x$.
Two small parameters are assumed:
$|\delta\op{\rho}(x)|\ll\rho_0$ (which makes this a quasicondensate), and 
$|\op{\theta}(x+\Delta x)-\op{\theta}(x)|\ll1$ 
(which is needed to ensure that the discretization of space corresponds to the continuum model). The latter is needed to self-consistently define the operator $\op{\theta}$  in \eqn{psiBog}. 
The exact quantum model is then truncated to 2nd or 3rd order in these small parameters, as the situation warrants, and 
the Hamiltonian takes the form 
\eq{HBog}{
\op{H} = \sum_{k\neq0} E_k \dagop{b}_k\op{b}_k + \frac{g}{2L}\op{P}^2 + E_{\rm ground}.
}
The $\op{b}_k$ ($\dagop{b}_k$) are quasiparticle annihilation (creation) operators for excited  plane wave modes, 
 with the usual commutation relation $[\op{b}_k,\dagop{b}_{k'}]=\delta_{kk'}$.
The system's description resembles an ideal gas of Bogoliubov quasiparticles. 
The quasiparticle energy 
is
\eq{Ek-M}{
E_k = \sqrt{\epsilon_k(\epsilon_k+2\mu)},
}
in terms of the free-particle energy $\epsilon_k = \frac{\hbar^2k^2}{2m}$ 
and chemical potential $\mu$.
The $k=0$ mode is represented by the background density $\rho_0$, while
$\op{P}$ is a dimensionless operator related to fluctuations in the total number of particles.
There is also an operator  
$\op{Q}$, which  is a zero energy collective coordinate for the global quantum phase. 
Both commute with all $\op{b}_k$ and $\dagop{b}_k$ and, moreover satisfy the relation
 $\left[\op{P},\op{Q}\right]=-i$.

To evaluate observables, 
the wavefunction elements in \eqn{psiBog}   can be expanded as: 
\eqa{wfeBog}{
\delta\op{\rho}(x) &=& \sqrt{\frac{\rho_0}{ L}}\sum_{k\neq0}(\wb{u}_k+\wb{v}_k)\left[e^{ikx}\op{b}_k + e^{-ikx}\dagop{b}_k\right] +\frac{\op{P}}{L}\nonu\\
\op{\theta}(x) &=& \frac{1}{2i\sqrt{\rho_0 L }}\sum_{k\neq0}(\wb{u}_k-\wb{v}_k)\left[e^{ikx}\op{b}_k - e^{-ikx}\dagop{b}_k\right] -\op{Q}\nonu\\
&&
}
where the quasiparticle wavefunction amplitudes are 
\eq{uvBog}{
\wb{u}_k\pm \wb{v}_k = \left[\frac{\epsilon_k}{\epsilon_k+2\mu}\right]^{\pm1/4}.
}

In thermal equilibrium all single operators $\op{b}_k$, $\dagop{b}_k$, $\op{P}$ and the anomalous average $\op{b}_k\op{b}_{k'}$ have zero mean, except for the occupations 
$\langle\dagop{b}_k\op{b}_{k'}\rangle = \delta_{kk'}n_k$, which are Bose-Einstein distributed:
\eq{nBog}{
n_k = \frac{1}{e^{E_k/k_BT}-1}.
}
Also, $\langle\op{P}{}^2\rangle=k_BT\,\frac{L}{g}$.
Averages involving  $\op{Q}$ are usually unnecessary. 

In the thermodynamic limit, the sum $\sum_{k\neq0}$ can be replaced\footnote{This is notwithstanding the fact that the theory in \cite{Mora03} is written in terms of a fine discretization of space with spacing $\Delta x$. This formally corresponds to $\int_{-\pi/\Delta x}^{\pi/\Delta x}$, but for any well described physical quantity the result must be unchanged in  the limit $\Delta x\to0$.}
 by $\frac{L}{2\pi}\int_{-\infty}^{\infty}dk$. 
Under the above  circumstances 
the equation of state 
is given by an integral:
\eqa{mu-M}{
n &=& \frac{\mu}{g} -\int_{-\infty}^{\infty} \frac{dk}{2\pi}  \left[(\wb{u}_k+\wb{v}_k)^2\,n_k + \wb{v}_k(\wb{u}_k+\wb{v}_k)\right].\qquad
}
This allows one to retroactively obtain $\mu/k_BT$ at given 
$\gamma$ and $\tau_d$ values.  They determine 
  $g$ and $n$ via \eqn{para}, apart from the one free scaling parameter $k_BT$.
Observables written in terms of $\op{\Psi}(x)$ can then be explicitly evaluated (see Appendix~\ref{S:BOGQ} for expressions).

%%%%%%%%%%%%%
\subsection{Classical field in the Bogoliubov regime}
\label{BOGC}
Let us now construct the classical field analogue for the extended Bogoliubov model.  
In general, a Bose field can be written in terms of mode functions $\psi_j(\bo{x})$ and mode annihilation operators $\op{a}_j$ as
\eq{bosefield}{
       \hat{\Psi}( {\bf x}) = \sum_{j} \hat{a}_j  \psi_j( {\bf x} ).
}
The underlying idea of classical field descriptions is that the creation/annihilation operators $\dagop{a}_j, \op{a}_j$ of highly occupied modes can be quite well approximated by complex amplitudes $\alpha_j\approx\op{a}_j$. This is because for a highly occupied mode with $\wb{n}_j=\langle\dagop{a}_j\op{a}_j\rangle$, the commutator $\left[\op{a}_j,\dagop{a}_j\right]\approx\mc{O}(1)$ is much smaller than $\sqrt{\wb{n}_j}$. 
The Bose field can then be approximated as 
\eq{cfield}{
       \hat{\Psi}( {\bf x}) 
       \to \Psi({\bf x}) = \sum_{j\in \mathcal{C}}  \alpha_j  \psi_j( {\bf x}).
}
For in-depth discussion of classical fields, we refer the reader to \cite{FINESS-Book-Davis,FINESS-Book-Cockburn,FINESS-Book-Brewczyk,FINESS-Book-Wright} and the earlier reviews \cite{Blakie08,Brewczyk07,Proukakis08}.

The c-field approximation corresponding to the extended Bogoliubov model 
is constructed using the quasiparticle modes $\op{b}_k$ in place of the $\op{a}_j$. 
In the uniform case these modes are plane waves with wavevector $k=2\pi j/L$. 
Several changes with respect to Sec.~\ref{BOGQ} need to be introduced to obtain the c-field description:
\begin{enumerate}
\item 
The approximation \eqn{cfield} breaks down for high energy modes, since they will be poorly occupied. For this reason, 
  the available set of modes should be restriced to a  subspace $\mc{C}$ specified by an energy cutoff $E_c$. In a uniform system, $E_c$ is equivalent to a certain cutoff wavevector $k_c$ such that $0<|k|\le k_c$. 
At the cutoff, the kinetic energy is $\ve_c=\hbar^2k_c^2/2m$, and in the particle-like regime above phonon excitations $E_c\approx\ve_c$.
Using the scaling of \eqn{para} with respect to the thermal de Broglie wavelength $\Lambda_T$, one can express $k_c$ in dimensionless form:
\eq{fc}{
       f_c = k_c \ \frac{\Lambda_T}{2\pi} = \frac{\hbar k_c}{\sqrt{2\pi mk_BT}}.
}   
\item The operators $\op{b}_k$ are replaced by appropriate random complex numbers $b_k$, that will give the required ensemble averages. The operator $\op{P}$ is replaced by random real values $P$ with variance $k_BT\,\frac{L}{g}$, which preserve its average, and $\op{Q}$ by a real phase $Q$ uniformly distributed on $[0,2\pi)$.
\item Quantum expectation values of operators $\langle\cdot\rangle$ are replaced by stochastic averages $\langle\cdot\rangle_s$ of c-field amplitudes. 
\end{enumerate}

The change from operators to c-numbers requires some care. Firstly, since we want to compare thermal equilibrium states,  we should
keep in mind that  c-fields equilibrate to Rayleigh-Jeans occupations, not Bose-Einstein. The correct thermal averages to use are then    $\langle b^*_kb_{k'}\rangle_s= \delta_{kk'}\,n^{\rm(cf)}_k$, with
\eq{nk-C}{
n^{\rm(cf)}_k = \frac{k_B T}{E_k}
}
and $\langle b_kb_{k'}\rangle_s =0$. Means $\langle b_k\rangle_s$ and $\langle P\rangle_s$ remain zero.

Secondly, now 
$b_k$ and $b_k^*$ commute, so that some observable expressions in thermal equilibrium need to be slightly modified. For example, $\langle b_kb_k^*\rangle_s = \langle b_k^*b_k\rangle_s = n^{\rm(cf)}_k$, in contrast to 
$\langle\op{b}_k\dagop{b}_k\rangle = \langle \dagop{b}_k\op{b}_k\rangle + 1$. 

In the thermodynamic limit, applying the above changes, \eqn{wfeBog} transform to:
\eqa{wfeCBog}{
\delta\op{\rho}(x) &\to& \frac{\sqrt{L\rho_0}}{2\pi}\int_{-k_c}^{k_c}dk(\wb{u}_k+\wb{v}_k)\left[e^{ikx}b_k + e^{-ikx}b^*_k\right] +\frac{P}{L}\nonu\\
\op{\theta}(x) &\to& \frac{\sqrt{L}}{4i\pi\sqrt{\rho_0}}\int_{-k_c}^{k_c}dk(\wb{u}_k-\wb{v}_k)\left[e^{ikx}b_k - e^{-ikx}b^*_k\right] -Q. \nonu\\
&&
}
The equation of state for classical fields becomes 
\eqa{mu-C}{
n &=& \frac{\mu^{\rm(cf)}(k_c)}{g} -\int_{-k_c}^{k_c} \frac{dk}{2\pi}  (\wb{u}_k+\wb{v}_k)^2\,n^{\rm(cf)}_k.
}
The chemical potential for a given density $n$ is different than the quantum one, and depends on the cutoff. That is, $\mu\to\mu^{\rm(cf)}(k_c)$ when evaluating $\wb{u}_k, \wb{v}_k, n^{\rm(cf)}_k$ or $E_k$.

Further quantities can be obtained using the same sequence of steps. as in \cite{Mora03}.
The local density-density correlation $g^{(2)}(z) = \tfrac{1}{n^2}\langle\dagop{\Psi}(x)\dagop{\Psi}(x+z)\op{\Psi}(x+z)\op{\Psi}(x)\rangle$ is expressed as 
\eq{g2-C}{
g^{(2)}_{\rm cf}(z) = 1+ \frac{2}{n}\int_{-k_c}^{k_c} \frac{dk}{2\pi} (\wb{u}_k+\wb{v}_k)^2n^{\rm(cf)}_k  \cos k z.
}

The coarse-grained density fluctuations in imaging bins measured in experiments \cite{Esteve06,Sanner10,Muller10,Jacqmin11,Armijo11,Armijo12} are
\eq{uGdef}{
u_G := \frac{{\rm var} \op{N}}{\langle \op{N} \rangle} 
}
so, substituting the Bogoliubov expressions, one obtains 
\eq{uG-C}{
u^{\rm cf}_G = 2\lim_{k\to0}(\wb{u}_k+\wb{v}_k)^2n^{\rm(cf)}_k = \frac{k_BT}{\mu^{\rm(cf)}(k_c)}.
}
The interaction energy per particle is trivially related to $g^{(2)}(0)$ straight from the Hamiltonian:
\eq{Eint-C}{
\mc{E}_{\rm int} = \frac{gn}{2}\,g^{(2)}_{\rm cf}(0)  =\mu^{\rm(cf)}(k_c)-k_BT \frac{\gamma}{4\pi \tau_d}.
}
The kinetic energy per particle is 
\eq{ve-C}{
\ve = \frac{\hbar^2}{2mn}\int_{-k_c}^{k_c} \frac{dk}{2\pi}\,k^2(1+2\wb{v}_k^2)n^{\rm(cf)}_k
}
and the total energy is $\mc{E}_{\rm tot} = \mc{E}_{\rm int}+\ve$.
Some additional expressions are given in Appendix.~\ref{S:BOGQ}.

%%%%%%%%%%%%%%%%%%%%%%%%%%%%%%%%%%%%%%%%%%%%%%%%%%%%%%%%%%%%%%%%
\section{Accuracy indicator}
\label{FOM-OPT}

The observables studied at hotter temperatures \cite{Pietraszewicz15,Pietraszewicz18a} were the local density fluctuations $g^{(2)}(z)$, coarse grained density fluctuations $u_G$ 
and the kinetic $\ve$, interaction $\mc{E}_{\rm int}$, and total $\mc{E}_{\rm tot}$ energies. 
 It was confirmed that three quantities ($u_G$, $\ve$, $\mc{E}_{\rm tot}$) 
suffice to produce a bound on the maximum deviation between c-field and exact predictions. 
 The discrepancies in all the other observables were consistently smaller.
Based on this observation, the maximum global error was defined as
\eq{RMSdef}{
	RMS(\gamma,\tau_d,f_c)= \sqrt{ \Big(\,\delta_{u_G}\,\Big)^2 +
       {\rm max}\Big[ \delta^{\,2}_{\ve}, \delta_{\mc{E}_{\rm tot}}^{\,2} \Big]  },
}
with the relative error $\delta_{\Omega}$ for a given observable $\Omega$.

We follow the same route here. The relative errors are
 \eq{delta}{
    \delta_{\Omega}=\Big[\ \frac{ \Omega^{\rm(Bog-cf)}(\gamma,\tau_d,f_c) }{\Omega^{\rm(Bog-q)}(\gamma,\tau_d) } -1 \Big],
}
where the fully quantum value is $\Omega^{\rm(Bog-q)}$ and the classical field value is $\Omega^{\rm(Bog-cf)}$. 
When comparing, we set the density $n$ in fully quantum and c-field results to be equal, so that they correspond to the same values of the $\gamma$ and $\tau_d$ parameters.

In the colder quasicondensate, we have checked for various parameter values 
that the three quantities used in \eqn{RMSdef} continue to have the largest errors compared to other observables.
(A representative case is shown in Fig.~\ref{sfig:rms-qqc} in Appendix~\ref{S:Observ}).
In this way we confirm that \eqn{RMSdef} is an adequate indicator of c-field accuracy in the entire quasicondensate regime.  

The minimum of $RMS$, min$RMS$ gives a figure of merit for the classical field description,
 and the value $f_c={\rm opt}f_c$ at which it occurs, gives the best cutoff to use. 
 Some analytic estimates are given in Sec.~\ref{S:BOGFC}.

%%%%%%%%%%%%%%%%%%%%%%%%%%%%%%%%%%%%%%%%%%%%%%%%%%%%%%%%%%%%%%%%%%%%%%%%%%%%%%%%%%%%%%%%%%%%%%%%%%%%%%%%%%%%%%%%%%%%%%%%%%%%%%%%%%%%%%%%%%%%%%%%%%%%%%%%%%%%%%%%%%%%%%%%%%%%%%%%%%%%%%%%%%%%%%%%%%%%%%%%%%%%%%%%%%%%%%%%%%%%%%%%%%%%%%%%%%%%
\section{The complete classical wave regime and optimal cutoff to use}
\label{RMS}

\begin{figure}[htb]
\begin{center}
\includegraphics[width=8cm]{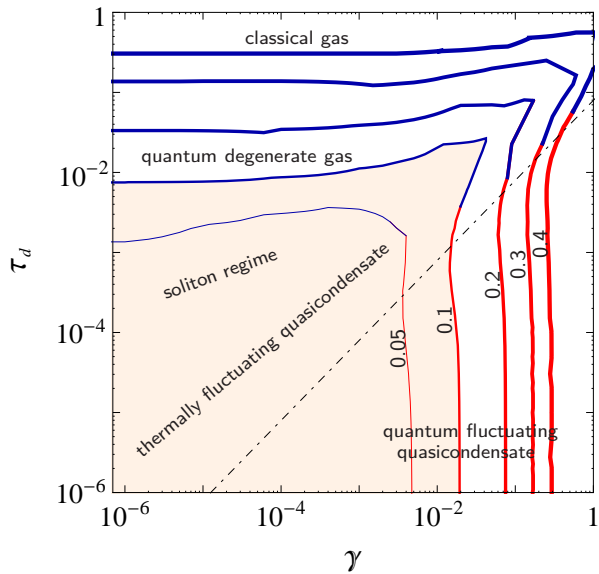}
\end{center}
\caption{
  The regime of applicability for classical fields, shown in light orange. In this region, observables are accurate to 10\% or better. 
 The values of the min$RMS$ indicator that bounds the accuracy are shown as a contour plot, with values printed on the figure.    
  The blue contours are from \cite{Pietraszewicz18a}, 
  while the red lines are obtained with the extended Bogoliubov theory used here.
  The dot-dashed line indicates the location of the $\mu\sim k_BT$ crossover between quasicondensates dominated by thermal and quantum fluctuations. 
 \label{fig:minRMS}}
\end{figure}
~
\begin{figure}[htb]
\begin{center}
\includegraphics[width=8cm]{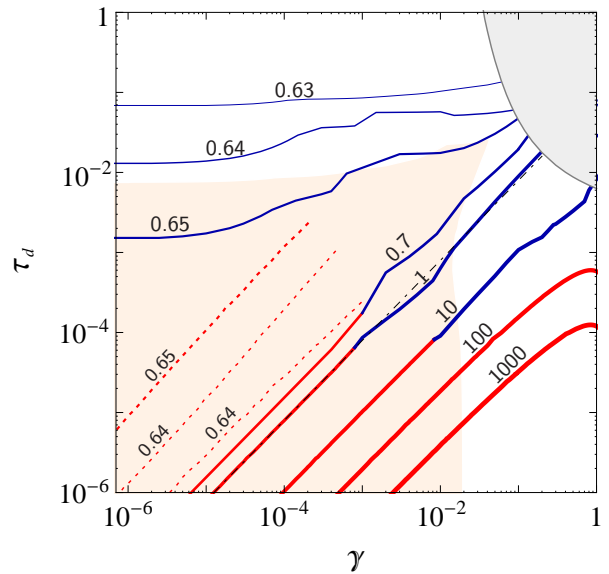}
\end{center}\vspace*{-4mm}
\caption{
 Globally optimal values of the cutoff opt$f_c$, shown with contours. 
 Notation as in Fig.~\ref{fig:minRMS}, and the light orange area indicates an accuracy of min$RMS<0.1$. 
 The gray colored area indicates a region from \cite{Pietraszewicz18a} 
 in which there was insufficient precision in the numerical ensembles to determine the position of the closely spaced contours.
\label{fig:optfc}}
\end{figure}

 We have calculated min$RMS$ and opt$f_c$ in the entire quasicondensate regime by evaluating the appropriate integrals for observables.
 This supplements the earlier results for $k_BT\gtrsim\mu$.
 Figs.~\ref{fig:minRMS}--\ref{fig:optfc} show a synthesis of these data sets, and are our main results.
 It is pleasing to note the perfect compatibility (contact) between the red contours obtained from the Bogoliubov theory
 and the blue contours obtained previously \cite{Pietraszewicz18a}.
Raw results are shown in Fig.~\ref{sfig:guts} in Appendix.~\ref{S:Observ}.

 Fig.~\ref{fig:minRMS} describes the \textbf{accuracy of the c-field description}. 
The light orange area in which accuracy is better than 10\% covers all temperatures from $\tau_d=0.008$ in the quantum degenerate region down to $T=0$, and covers the whole dilute gas up to around $\gamma=0.018$. It also extends somewhat further  up to
around $\tau_d\approx \gamma\approx0.03$, for reasons that are not  understood at the moment.

It can now be seen that all the low order observables remain well described in the quantum fluctuating region, even down to $T=0$. This is something that
 was not obvious \emph{a priori} since
the weak antibunching that occurs 
 due to quantum depletion
 ($g^{(2)}(0)\approx 1 - 2\sqrt{\gamma}/\pi$ \cite{Sykes08,Deuar09}) cannot be correctly
 replicated by classical fields. 
 However, this has little effect on the coarse-grained density fluctuation statistics 
 \eq{uG}{
 u_G = 
  n\, \int d{\bf z}\, {\big[} g^{\rm (2)}({\bf z})-1 {\big]} + 1.
 }
 The reason is that the two contributions to $u_G$
 that are missing in classical fields (shot noise ``+1'' and antibunching in $g^{(2)}$) cancel in the full quantum description. 
 In the c-field description, 
$g^{(2)}_{\rm  cf}(z) = \langle |\Psi(x)|^2|\Psi(x+z)|^2\rangle_s/n^2$,  
and from the definition \eqn{uGdef} the number fluctuations are related by
 \eq{uGcf}{
   u_G^{\rm cf} :=  n \int d{\bf z}\, {\big[} g^{\rm (2)}_{\rm cf}({\bf z})-1 {\big]}.
 }
instead of \eqn{uG}.
The shot noise ``+1'' term is no longer present, and values of $u_G$ tending to zero can continue to be obtained despite $g^{(2)}_{\rm cf}(z)>1$.

 In turn, the dependence of the \textbf{optimal cutoff} opt$f_c$ is shown in Fig.~\ref{fig:optfc}. 
The main standout feature is that there are different behaviors depending on whether $k_BT\gtrsim\mu$ or $k_BT\lesssim\mu$, with a changeover marked by the dot-dashed line near the opt$f_c=1$ contour. 

 In the thermal upper part of the diagram studied already in \cite{Pietraszewicz18a}, one has a practically constant
 ${\rm opt}f_c=0.64\pm0.01$, indicating that one cutoff choice is appropriate for the whole cloud when $k_BT\gtrsim\mu$. 
 The Bogoliubov data (in red) also show this  but are more precise at low temperatures.
 They indicate the presence of a broad shallow trough, between dashed red lines with values opt$f_c$=0.64 in Fig.~\ref{fig:optfc}. 
 The trough must disappear at higher temperatures since it was not seen in \cite{Pietraszewicz18a}, while
the correctness of the Bogoliubov decreases with growing temperature.

 The lower part of Fig.~\ref{fig:optfc} confirms the conjecture voiced in \cite{Pietraszewicz18a} that 
 a change of cutoff behavior begins when quantum fluctuations dominate. A rapid growth of opt$f_c$ is observed, 
 and approximated by
 \eq{optfcBog}{
 {\rm opt}f_c = \frac{1}{12\pi^2}\left(\frac{\gamma}{\tau_d}\right)^{\frac{3}{2}}\left[ 
 1 -\frac{7\sqrt{\gamma}}{2\pi} + \frac{3\pi^2\tau_d}{\gamma} + \dots\right].
 }
 See Sec.~\ref{S:BOGRMS}. Concurrently, 
 \eq{minRMSBog}{
 {\rm min}RMS = \frac{\sqrt{5\gamma}}{\pi}\left[1 +\frac{\sqrt{\gamma}}{30\pi}+\pi^2\frac{\tau_d}{\gamma} + \dots 
 \right].
 }

\begin{figure}[htb]
\begin{center}
\includegraphics[width=0.49\columnwidth]{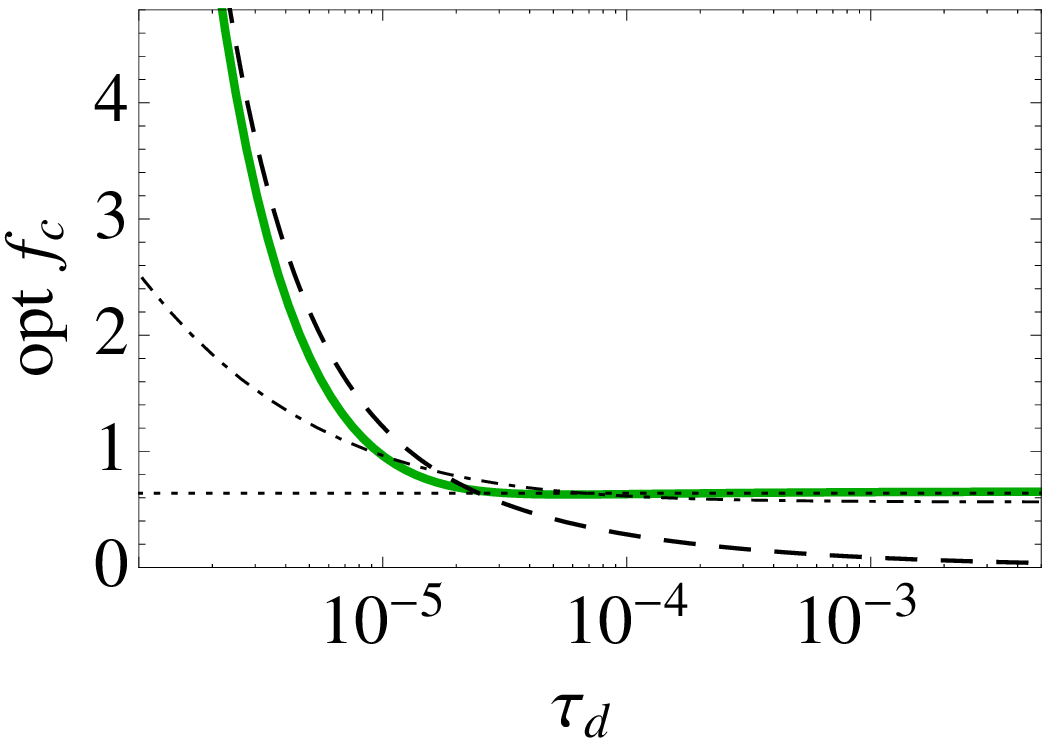}
\includegraphics[width=0.49\columnwidth]{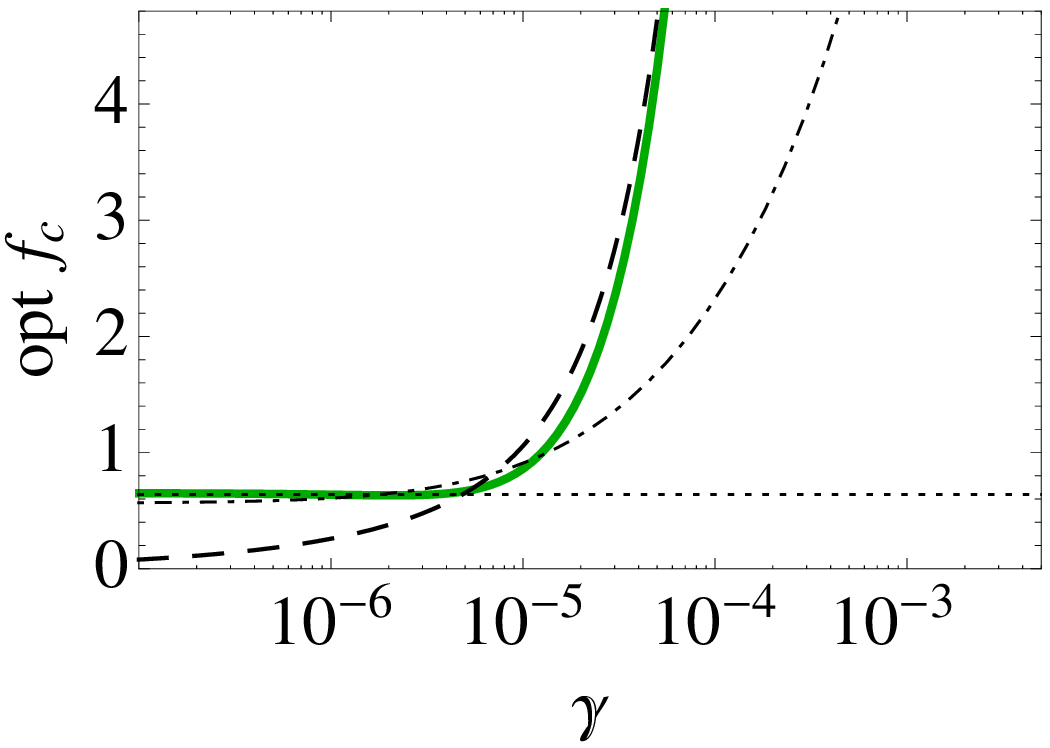}
\end{center}\vspace*{-4mm}
\caption{
Optimal cutoff shown for two characteristic slices in parameter space (in green). 
Left panel:  $\gamma=1.2 \times 10^{-4}$. Right panel: $\tau=10^{-6}$.
The black lines show the two approximations:  
 opt$f_c=0.64$ -- (dotted), 
\eqn{optfcBog} -- (dashed), as well as
the hitherto ``rule of thumb'' \eqn{fcpres} (dot-dashed).
\label{fig:slices}}
\end{figure}

The dependence $\gamma/\tau_d \propto gn/T$ means that the cutoff begins to strongly depend on density in this region. 
This suggests that a larger range of momenta $k$ should be allowed in the center of the cloud than in the tails. 
A plane-wave basis does not provide such a possibility, but a harmonic oscillator basis does, as studied in \cite{Bradley05}. 
Hence, unlike at higher temperatures, clouds whose central region reaches $\mu\gtrsim k_BT$ should use bases that take into account the trap shape.

%%%%%%%%%%%%%%%%%%%%%%%%%%%%%%%%%%%%%%%%%%%%%%%%%%%%%%%%%%%%%%%%%%%%%%%%%%%%%%%%%
\section{Kinetic energy and previous cutoff determinations}
\label{ruleTh}

 The reason why a high cutoff is needed in the low temperature quasicondensate 
 is that the kinetic energy begins to rise steeply with $\gamma$.
It can be shown that $\ve \approx \frac{k_BT}{6\pi^2}\,\frac{\gamma^{3/2}}{\tau_d}$  there (see App.~\ref{S:BOGFC}).
 
The kinetic energy is contained in repulsive quantum fluctuations.
In a c-field description quantum fluctuations are absent, so  
to build up the correct level of kinetic energy, extra modes (with $k_BT$ energy in each) should be introduced (details in App.~\ref{S:LOWT}).
Adding these extra modes does not adversely affect other important observables since their occupations are small.

The main cutoff result \eqn{optfcBog} can be compared to the widely used ``rule of thumb'' \cite{Blakie08,Bienias11a,FINESS-Book-Wright,Kashurnikov01}. This rule of thumb says that
the single particle energy at the cutoff should be $\approx k_BT+\mu$, and for a plane-wave basis this energy is $\ve_c=\pi (f_c)^2 k_BT$.  Using \eqn{muT} and \eqn{fc} leads to:
\eq{fcpres}{
f_c^{\rm thumb} \approx \sqrt{\frac{1}{\pi}\left(\frac{\gamma}{2\pi\tau_d} + 1\right)}.
}
 Note that both \eqn{fcpres} and \eqn{optfcBog} grow with the ratio $\gamma/\tau_d$, but the 
 global opt$f_c$ \eqn{optfcBog} grows with a faster power law. The difference can be seen in Fig.~\ref{fig:slices}.  

It is informative to look at the length scales allowed  by the two cutoffs. 
The smallest length scale accessible with a $k_c$ cutoff in a plane wave basis is about $\pi/k_c = \Lambda_T/2f_c$.
For $k_BT\gtrsim\mu$ the accessible length scales reached up to the thermal de Broglie wavelength. 
However, for $k_BT\lesssim\mu$, one should resolve the healing length
$\xi = \frac{\hbar}{\sqrt{m\mu}}$. The rule of thumb cutoff \eqn{fcpres} leads to  $\pi/k_c
\approx~\frac{\pi}{\sqrt{2}}\xi$, 
so that this resolution is achieved. 
In turn, the optimum cutoff \eqn{optfcBog} allows smaller length scales down to $\pi/k_c\approx 3\pi(\frac{k_BT}{\mu})\xi$. 

\begin{figure}[h!]
\begin{center}
\includegraphics[width=0.9\columnwidth]{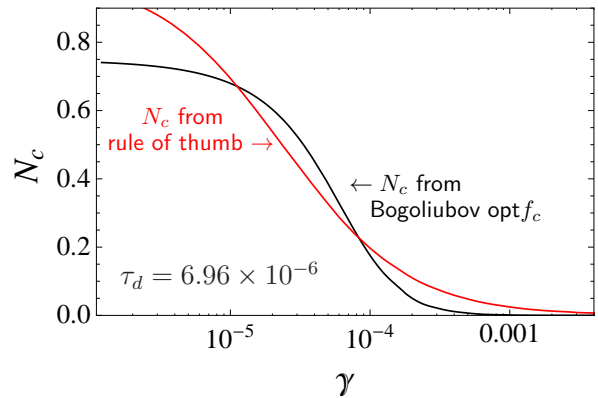}
\end{center}\vspace*{-4mm}
\caption{
Occupation of the highest energy mode $N_c$ when the cutoff is given by  opt$f_c$ (black line) and the rule of thumb \eqn{fcpres} (red line). The asymptotic value at low $\gamma$ is 0.746.
\label{fig:Nc}}
\end{figure}

In the history of the field, the cutoff has also been characterized by the c-field occupation $N_c$ of the (quasi)particle mode with the highest energy \cite{Davis01a,Zawitkowski04,Blakie08}. It is expected that $N_c\sim1$ from general arguments. 
In a Bogoliubov quasiparticle treatment one has
\eq{Nbog}{
N_c \approx \frac{k_BT}{\sqrt{\ve_c(\ve_c+2\mu)}}.
}
Fig.~\ref{fig:Nc} presents the value corresponding to the numerically calculated  opt$f_c$. 
Its analytic estimate for $k_BT\ll\mu$ is 
\eq{NclowT}{
N_c = 18 \left(\frac{2\pi\tau_d}{\gamma}\right)^3\left[ 1 +\frac{7\sqrt{\gamma}}{\pi}-\frac{6\pi^2\tau_d}{\gamma} + \dots 
\right]
\approx
18 \left(\frac{k_BT}{\mu}\right)^3.\quad
}
At small $\gamma$, $N_c$ takes the value 0.746. 
It initially looks surprising that $N_c$ plummets to zero in the quantum fluctuating regime.
However, the rule of thumb \eqn{fcpres} also predicts a rapidly falling $N_c$ behavior with $\gamma$ (the red line in Fig.\ref{fig:Nc}), only that the fall is less steep: $N_c\approx \frac{1}{\sqrt{3}}\left(k_BT/\mu\right)$. One can see that these almost empty modes are needed to allow physically important length scales and correct kinetic energy. 
Such a relaxation of the usual criterion of $\mc{O}(1)$ cutoff mode occupation has also precedents in the truncated Wigner prescription \cite{Norrie06}.

%%%%%%%%%%%%%%%%%%%%%%%%%%%%%%%%%%%%%%%%%%%%%%%%%%%%%%%%%%%%%%%%%%%%%%%%%%%%%%%%%%%%%%%%%%%%%%%%%%%%%%%%%%%%%%%%%%%%%%%%%%%%%%%%%%%%%%%%%%%%%%%%%%%%%%%%%%%%%%%%%%%%%%%%%%%%%%%%%%%%%%%%%%%%%%%%%%%%%%%%%%%%%%%%%%%%%%%%%%%%%%%%%%%%%%%%%%%%
\section{Conclusions}
\label{CONCLUSIONS}

The effectiveness of the classical field description has now been assessed across the whole 1d Bose gas, completing the campaign started in \cite{Pietraszewicz15,Pietraszewicz18a}.   
Figs.~\ref{fig:minRMS}-~\ref{fig:optfc} are a synthesis of the results: the first shows the accuracy that is possible in many observables simultaneously, while the second figure specifies the cutoff that achieves this.
The light orange region specifies the parameters for which accuracy is within 10\% or better, and the dominant physics is indeed that of matter waves. 
Simulations can be confidently carried out provided the system stays in this region. Conversely, outside of this region, one or more of the standard observables are always going to be inaccurate.

Basically, there are two main regions  of interest. 
The first is  $k_BT\gtrsim\mu$, characterized by an optimal cutoff opt$f_c=0.64$ that depends only on temperature \cite{Pietraszewicz18a}. 
This result should be applicable to nonuniform gases even in a plane wave basis 
and covers the thermal quasicondensate, the soliton regime, and most of the degenerate gas.  

The second region is
the $k_BT\lesssim\mu\approx gn$ quasicondensate regime, studied here, in which quantum fluctuations are important.
The optimal cutoff in this regime is opt$f_c\approx\frac{1}{3\sqrt{2\pi}}(\mu/k_BT)^{3/2}$. It lies at energies well above $k_BT$ and becomes strongly density dependent.
 This high cutoff is needed to correctly capture the kinetic energy held in quantum fluctuations.
Importantly, it does not distort most observables, because the occupation of the additional modes is very low ($N_c\approx 18 (\frac{k_BT}{\mu})^3$). This goes against the 
common intuition that the  cutoff mode occupation should be of $\mc{O}(1)$.

 A high cutoff here is actually a welcome result because studies of defect evolution at low temperature use
 very high resolution numerical grids that have an energy cutoff well above $k_BT$, and would be suspect if $N_c\sim\mc{O}(1)$ was required.  
 We can also conclude that a plane wave basis will not be accurate for nonuniform clouds
 whose central density exceeds $gn\gtrsim k_BT$.  Bases that take into account the trap shape are then needed. 

Looking forward, we have seen that the c-field variant of the extended Bogoliubov model described in Sec.~\ref{BOGC} allows one to easily reach the low temperature limit. 
It can also be used to investigate the case of 2d and 3d gases, which may behave very differently.

%%%%%%%%%%%%%%%%%%%%%%%%%%%%%%%%%%%%%%%%%%%%%%%%%%%%%%%%%%%%%%%%%%%%%
%Acknowledgments
%%%%%%%%%%%%%%%%%%%%%%%%%%%%%%%%%%%%%%%%%%%%%%%%%%%%%%%%%%%%%%%%%%%%%
\acknowledgments
This work was supported by the National Science Centre (Poland) grant No. 2012/07/E/ST2/01389. 

\bibliography{cfields}

\appendix

%%%%%%%%%%%%%
\section{Accuracy of the extended Bogoliubov theory }
\label{S:BOGYY}

\begin{figure}[htb]
\begin{center}
\includegraphics[width=0.8\columnwidth]{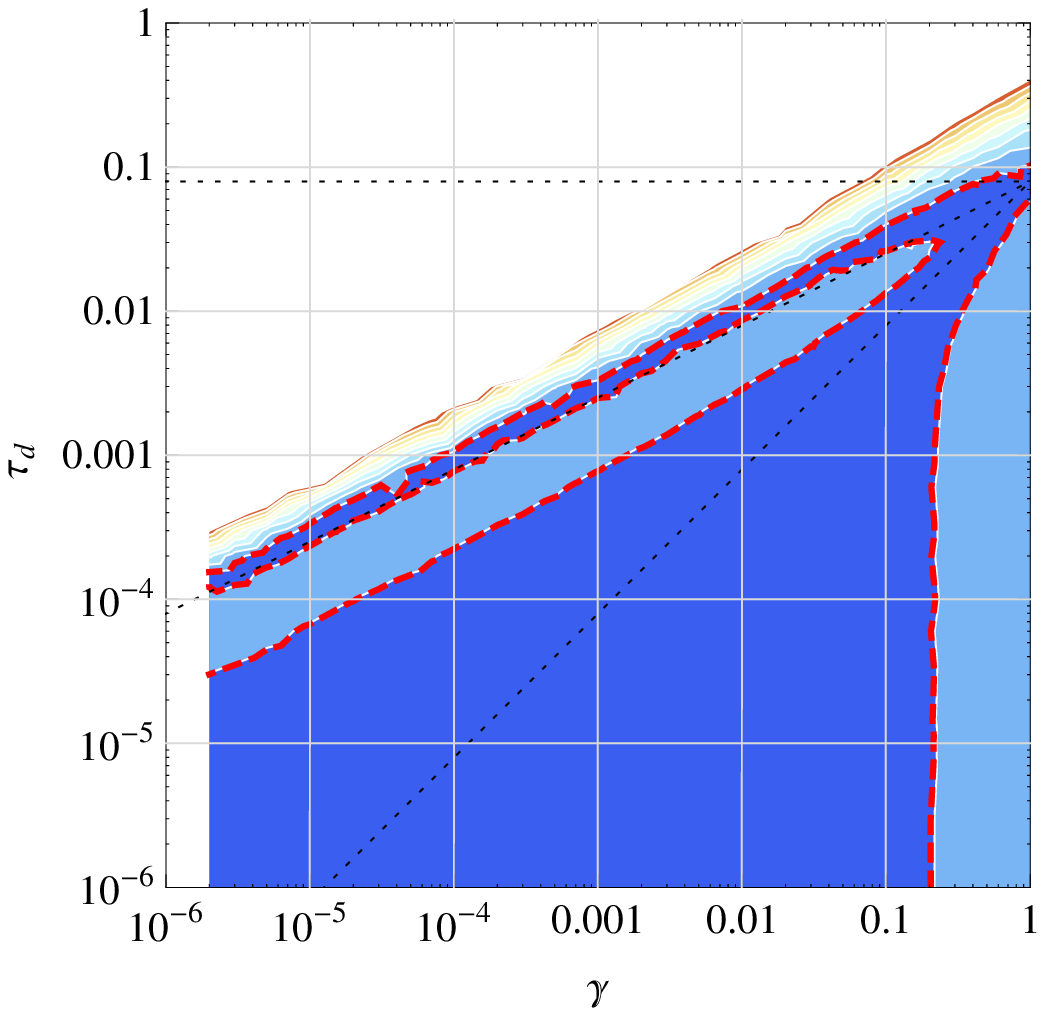}
\end{center}\vspace*{-4mm}
\caption{
This diagram shows the accuracy of the extended Bogoliubov treatment with respect to exact Yang-Yang theory by plotting contours (at 0.1,0.2,\dots,1.0) of \eqn{RMSYY}.
The dark blue region together with a thick red dashed line indicates the $RMS^{(Q)}\leq 0.1$ contour. 
\label{sfig:bogyy}}
\end{figure}

A figure of merit for the accuracy of the Bogoliubov theory can be defined in a similar way to \eqn{RMSdef}, by comparing to the exact quantum solution:
\eq{RMSYY}{
	RMS^{(Q)}(\gamma,\tau_d)= \sqrt{ (\delta^{(\rm Bog)}_{u_G})^2 + 
       {\rm max}\left[ 
 (\delta^{(\rm Bog)}_{\ve})^2,
(\delta^{(\rm Bog)}_{\mc{E}_{\rm tot}})^2
\right]  },
}
where
\eq{rmsYY}{
     \delta^{(\rm Bog)}_{\Omega}(\gamma,\tau_d) =  
     \Bigg(  \frac{ \Omega^{\rm (Bog-q) }(\gamma,\tau_d) }{ \Omega^{\rm(q)}(\gamma,\tau_d)} - 1 \Bigg).
}
The accurate region with less than 10\% error is shown in Fig.~\ref{sfig:bogyy} in dark blue, circumscribed by the red dashed line.
We restricted our use of Bogoliubov data in the synthesis of Figs.~\ref{fig:minRMS} and~\ref{fig:optfc} to far within this accurate region.

%%%%%%%%%%%%%%%%%%%%%%%%%%%%%%
\section{Calculations with extended Bogoliubov}
\label{S:BOG}

%%%%%%%%%%%%%
\subsection{Observable expressions}
\label{S:BOGQ}
Following on from Sec.~\ref{BOGQ},
the fully quantum expressions for the observables are \cite{Mora03}:
\eq{g2-M}{
g^{(2)}(z) = 1+ \frac{2}{n}\int_{-\infty}^{\infty} \frac{dk}{2\pi} \left[(\wb{u}_k+\wb{v}_k)^2n_k + \wb{v}_k(\wb{u}_k+\wb{v}_k)\right] \cos k z.
}
leading via \eqn{uG} to 
\eq{uG-M}{
u_G = 1+2\lim_{k\to0}\left[(\wb{u}_k+\wb{v}_k)^2n_k+\wb{v}_k(\wb{u}_k+\wb{v}_k)\right] = \frac{k_BT}{\mu}.
}
This is a convenient form like \eqn{uG-C}. The interaction energy per particle continues to be related to $g^{(2)}(0)$ through \eqn{Eint-C}, giving 
\eq{Eint-M}{
\mc{E}_{\rm int} =\mu-k_BT \frac{\gamma}{4\pi \tau_d}.
}
The kinetic energy per particle is 
\eq{ve-M}{
\ve = \frac{\hbar^2}{2mn}\int_{-\infty}^{\infty}\!\frac{dk}{2\pi}\,k^2\left[(1+2\wb{v}_k^2)n_k + \wb{v}_k^2\right].
}
Phase correlations are
\eq{g1-M}{
g^{(1)}(z) = \exp\left[  -\frac{1}{n}\int_{-\infty}^{\infty} \frac{dk}{2\pi}\left[(\wb{u}_k+\wb{v}_k)^2n_k + \wb{v}_k^2 \right](1-\cos k z)\!\right].
}
For the corresponding c-field description, one has the slightly modified expression
\eq{g1-C}{
g^{(1)}_{\rm cf}(z) = \exp\left[ -\frac{1}{n}\int_{-k_c}^{k_c} \frac{dk}{2\pi}(\wb{u}_k+\wb{v}_k)^2n^{\rm(cf)}_k(1-\cos k z)\right].
}

Consideration was also given to the condensate mode occupation, $N_0$, i.e. the number of atoms in the $k=0$ mode. This observable tends to a well defined constant value as the gas length grows, but of course it becomes negligible compared to $N$ in the thermodynamic limit of the 1d gas.  It is important for comparison to earlier cutoff determinations made in low temperature mid-size systems with $\mc{O}(1000)$ atoms, where condensate fraction $n_0=N_0/N$ remained significant \cite{Witkowska09,Bienias11a}. 
Since the density in k space can be expressed as $\wt{n}(k) = \frac{N}{2\pi}\int_{-\infty}^{\infty} dz\, g^{(1)}(z)\, e^{ikz}$, \eqn{g1-M} can be used to obtain the occupation of the lowest energy ($k\to0$) state:
\eq{N0-Q}{
N_0 = \wt{n}(0)\Delta k = n\int_{-\infty}^{\infty} dz\, g^{(1)}(z).
} 
For the c-field description,  $N_0$ continues to be given by the form \eqn{N0-Q} but now using $g^{(1)}_{\rm cf}(z)$ from \eqn{g1-C}.

%%%%%%%%%%%%%%%%%%%%%%%%%%%%%%
\subsection{Relative errors and optimization}
\label{S:Observ}

\begin{figure}
\begin{center}
\includegraphics[width=0.49\columnwidth]{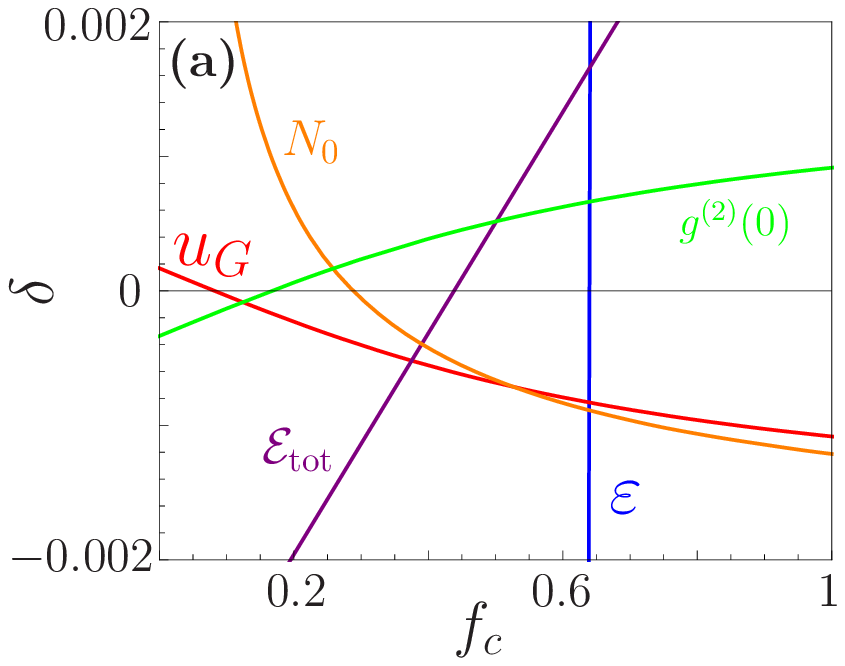}
\includegraphics[width=0.49\columnwidth]{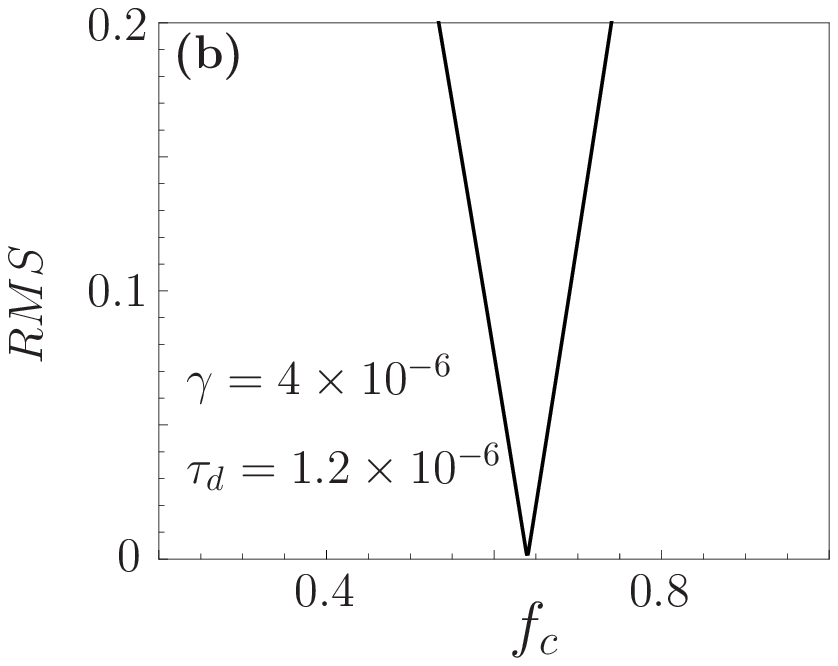}
\includegraphics[width=0.49\columnwidth]{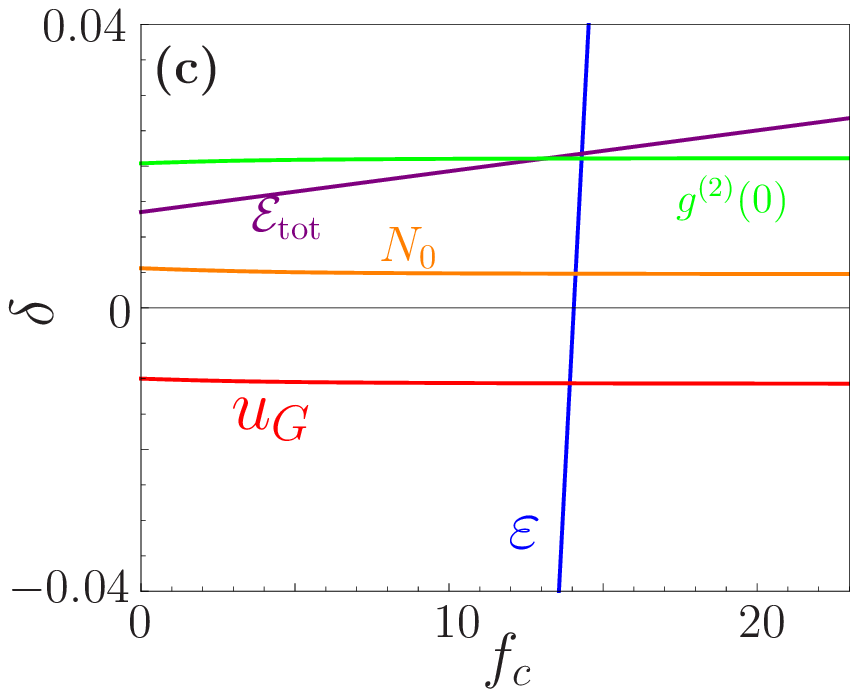}
\includegraphics[width=0.49\columnwidth]{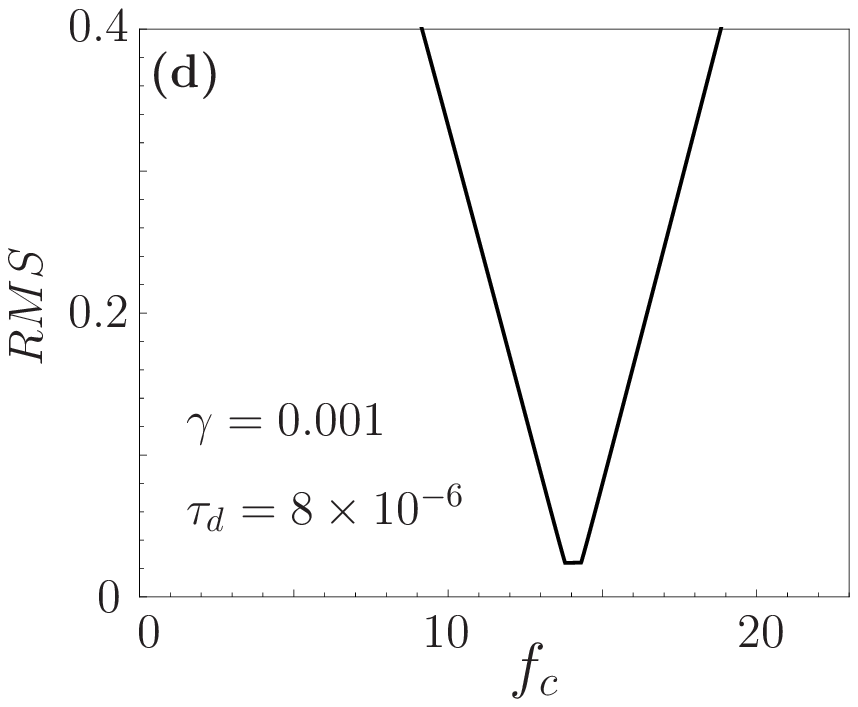}
\end{center}
\caption{
Cutoff dependence in the quasicondensate. The top panels are in the thermally dominated regime $\gamma=4\times 10^{-6}$, $\tau_d=1.2\times 10^{-6}$, and the lower panels in the quantum fluctuation regime $\gamma= 10^{-3}$, $\tau_d=8\times 10^{-6}$.
Notation: $g^{(2)}(0)$ (green), $u_G$ (red), $\mc{E}_{\rm tot}$ (purple), $\ve$ (blue), $N_0$ (orange). 
Left panels: discrepancies for single observables calculated with \eqn{delta};  Right panels: global discrepancy $RMS$ calculated with \eqn{RMSdef}. 
}
\label{sfig:rms-qqc}
\end{figure}

 The cutoff-dependent discrepancy of various observables in the quasicondensate regimes are shown in Fig.~\ref{sfig:rms-qqc}, calculated using \eqn{delta}.
 We observe that $\ve$ has the the most extreme rising behavior,
 while $u_G$ captures the strongest
 falling behavior in the vicinity where all errors are small.
 Hence, the best cutoff occurs at a point where there is a balance between these rising and falling predictions. 
The goodness of the classical field description depends on how large the actual discrepancies at this point are.
The maximal error at such an optimal cutoff can be set by $\mc{E}_{\rm tot}$ or $u_G$. 

\begin{figure}[htb]
\begin{center}
\includegraphics[width=0.8\columnwidth]{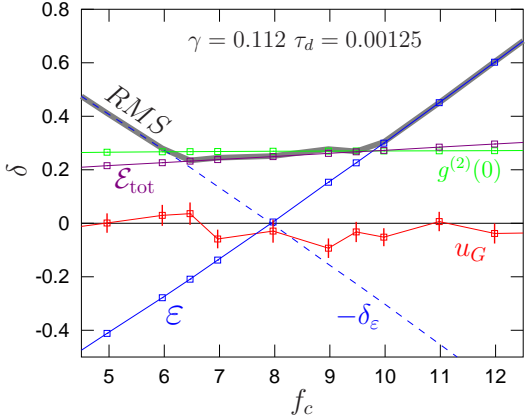}
\end{center}\vspace*{-4mm}
\caption{
Cutoff dependence of the discrepancies $\delta_{\Omega}$
for the case $\gamma = 0.112,\tau_d = 0.00125$ (from \cite{Pietraszewicz18a}).
Notation as in Fig.~\ref{sfig:rms-qqc}.
The figure of merit $RMS(f_c)$ is shown as the thick grey line.
The dashed blue line shows $-\delta_{\ve}$ as a  reference.
}
\label{sfig:rms}
\end{figure}

The behavior in the quantum fluctuating condensate is somewhat similar to the behavior seen at large $\gamma$ in \cite{Pietraszewicz18a}, reproduced here in Fig.~\ref{sfig:rms}. 
This figure also shows  a flat-bottomed minimum in $RMS$ more clearly than in Fig.~\ref{sfig:rms-qqc}d because it is much broader when $\gamma$ is large.
For more details see Sec.~\ref{S:BOGFC}.

\begin{figure*}[htb]
\begin{center}
\includegraphics[width=0.95\columnwidth]{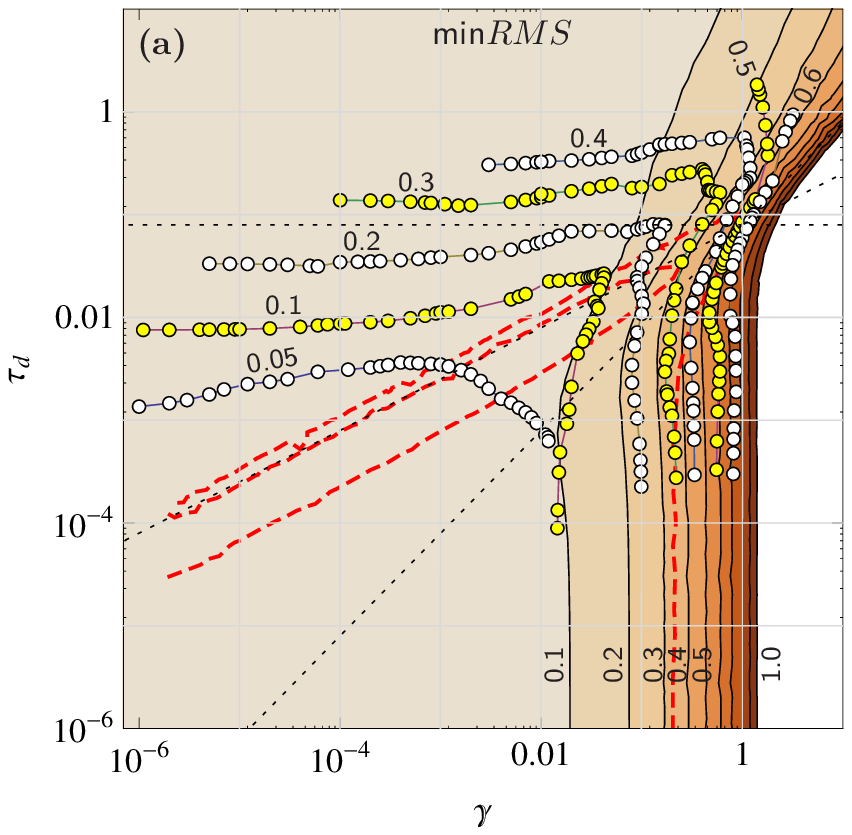}
\hspace*{0.08\columnwidth} 
\includegraphics[width=0.95\columnwidth]{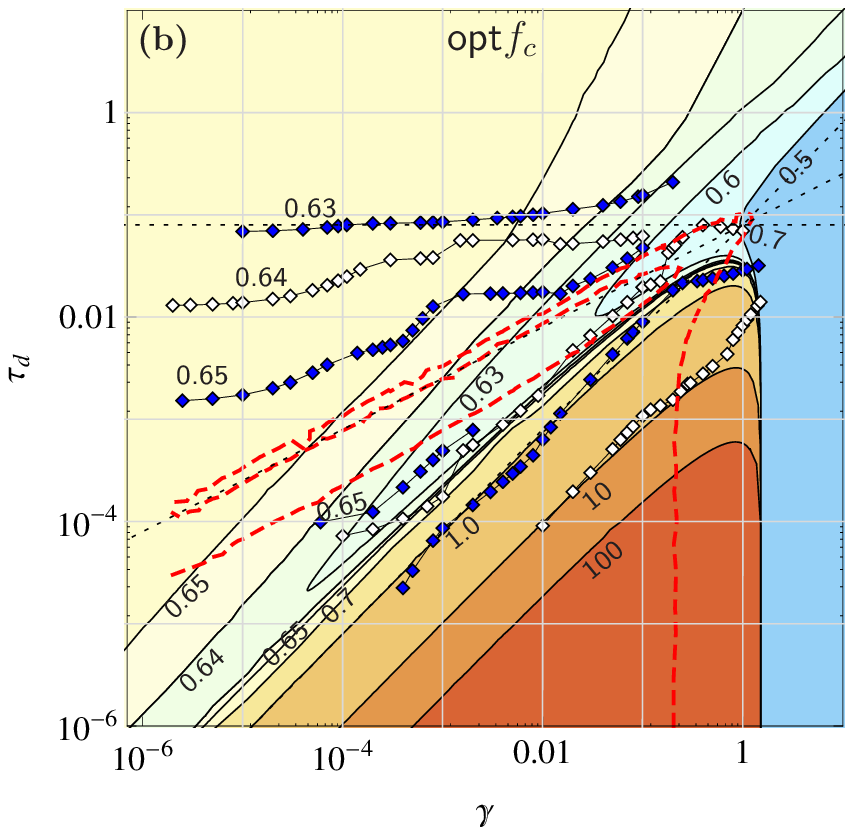}
\end{center}\vspace*{-4mm}
\caption{
A detailed comparison of raw Bogoliubov and numerical ensemble results. Contours of min$RMS$ (a) and opt$f_c$ (b) are shown with numbered values on the plot, similarly to Figs.~\ref{fig:minRMS} and~\ref{fig:optfc}, respectively. Bogoliubov results are shown as solid contours between colored fields,
 while contours derived from the numerical ensembles \cite{Pietraszewicz18a} are shown as joined symbols.
 The red dashed line is copied from Fig.~\ref{sfig:bogyy} and shows the location at which the error between the Bogoliubov and exact results reaches 10\%.
\label{sfig:guts}}
\end{figure*}

In Fig.~\ref{sfig:guts} one can see the results for min$RMS$ and ot$f_c$ obtained from the Bogoliubov calculations here,
 overlaid with the numerical ensemble results from \cite{Pietraszewicz18a}.
 The data analysis procedure was described in detail in \cite{Pietraszewicz18a}. 
 For the Bogoliubov case, the data is finely spaced and smooth, and a simple application of
 the the Wolfram \textsl{Mathematica} algorithm \texttt{ContourPlot} turned out to be sufficient for the task, without the need for Lagrangian interpolation.

%%%%%%%%%%%%%%%%%%%%%%%%%%%%%%
\section{Kinetic energy and cutoff in the quantum fluctuation region }
\label{S:LOWT}

\begin{figure}[b]
\begin{center}
\includegraphics[width=0.8\columnwidth]{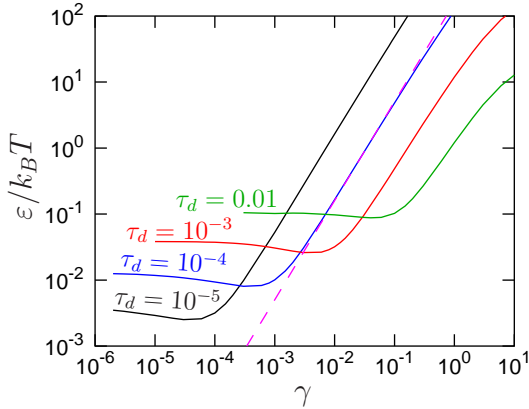}
\end{center}\vspace*{-4mm}
\caption{
Kinetic energy $\ve$ as a function of $\gamma$. Exact Yang-Yang results.
The magenta dashed line shows the approximation \eqn{ve-M-2-est} for $\tau_d=10^{-4}$.
}
\label{sfig:ekin}
\end{figure}

The exact Yang-Yang results for kinetic energy $\ve$ are shown in Fig.~\ref{sfig:ekin} and display a rapid increase once the regime $\gamma\gtrsim2\pi\tau_d$ is reached.
The reason for this rapid growth can be tracked to the kinetic energy present within the quantum fluctuations. 
To see this, let us make some  approximations to \eqn{ve-M} in the $\mu\gg k_BT$ regime.

First, consider the thermal part that contains $n_k$. 
The main contributing modes are in the phonon regime where  $E_k \approx \sqrt{2\ve\mu}=\xi |k|\mu$, 
and their occupation 
can be approximated by $n_k\approx k_BT/E_k$. 
The quantity $\wb{v}_k^2\approx1/2\xi|k|$ is much greater than one in this regime.
Using all this, the thermal term in \eqn{ve-M} can be written as 
\eqa{ve-M-1-est}{
\ve_{\rm th} &\approx& \frac{\hbar^2}{2mn}\int_{-k_BT/\mu\xi}^{k_BT/\mu\xi}\ \frac{dk}{2\pi} k^2 \frac{1}{\xi|k|}\, \frac{k_BT}{\xi|k|\mu}
= \left(\frac{k_BT}{\mu}\right)^2 \frac{\mu}{2\pi\xi n} \nonu\\
&\approx& \frac{k_BT\tau_d}{\sqrt{\gamma}} 
}

 The other part of  \eqn{ve-M} (the quantum fluctuation term) contains just $\wb{v}^2_k$. 
 This quantity decays rapidly $\wb{v}_k^2\approx1/(k\xi)^4$ in the particle regime, and has negligible contribution there.
 Hence, just the phonon contribution is relevant.
 The crossover in the behavior of $\wb{v}_k$ to particle-like is at $|k|=2^{1/3}/\xi$, so the quantum fluctuation part is 
\eq{ve-M-2-est}{
\ve_{\rm qf} \approx \frac{\hbar^2}{2mn}\int_{-2^{1/3}/\xi}^{2^{1/3}/\xi}\ \frac{dk}{2\pi} k^2 \frac{1}{2\xi|k|}
= \frac{\mu}{2^{5/3}\pi n\xi}
\approx \frac{k_BT\gamma^{3/2}}{4\pi^22^{2/3}\tau_d} 
}
 This is far larger than \eqn{ve-M-1-est}, and indicates that kinetic energy is indeed dominated by quantum fluctuations.

 Let us now see what happens in the c-field description. The expression \eqn{ve-C} contains only a thermal term,
 but $n^{\rm(cf)}_k=k_BT/E_k$ does not decay as fast as in a Bose-Einstein distribution and both  phonon-like and particle-like modes contribute.
 Using again the crudest useful approximation, the phonon and particle regimes meet at $|k|=2/\xi$. Taking the leading terms, 
\eqa{ve-C-est}{
\ve^{\rm(cf)} &\approx& \frac{\hbar^2}{mn}\left\{ \int_0^{2/\xi}\frac{dk}{2\pi} k^2 \frac{1}{\xi|k|} \frac{k_BT}{\xi|k|\mu} + \int_{2/\xi}^{k_c}\frac{dk}{2\pi} k^2 \frac{2k_BT}{\mu(\xi k)^2}\right\}
\nonu\\
& = & \frac{k_BT k_c}{\pi n}\left(1-\frac{1}{f_c}\sqrt{\frac{\mu}{2\pi k_BT}}\right)\nonu\\
& \approx & \frac{k_BT k_c}{\pi n} = 2 k_BT f_c \sqrt{\tau_d}.
}
 The second term in the brackets on the 2nd line turns out to be small once the  estimate \eqn{fc-estmp} is obtained.
 The estimate \eqn{ve-C-est} corresponds to assuming exactly $k_BT$ purely \emph{kinetic} energy per mode.
 Since the vast majority of modes are particle-like because of the high cutoff, this is actually a reasonable approximation. 

\eqn{ve-C-est} can be compared to the quantum kinetic energy \eqn{ve-M-2-est}. Such a comparison gives the following prediction for the cutoff based on kinetic energy alone:
\eq{fc-estmp}{
f_c \approx 0.008\left(\frac{\gamma}{\tau_d}\right)^{\frac{3}{2}}.
}
\eqn{fc-estmp} agrees remarkably well with the Bogoliubov result \eqn{optfcBog} and the exact numerics.	

%%%%%%%%%%%%%%%%%%%%%%%%%%%%%%
\section{Bogoliubov estimates for \lowercase{opt$f_c$} and \lowercase{min}$RMS$ in the quantum fluctuating regime}

\label{S:BOGFC}

The quantum fluctuating Bogoliubov regime has two small parameters: $\gamma\ll1$ and a temperature scaled with respect to the chemical potential:
\eq{smallt}{
t = \frac{2\pi\tau_d}{\gamma} \approx \left(\frac{k_BT}{\mu}=u_G\right)\ll1.
}
The equality $k_BT/\mu=u_G$ follows from  \eqn{uG-M} and \eqn{uG-C}.
We will make a self-consistent expansion of the required quantities in these small parameters. 
We know from \eqn{fc-estmp} that the scaling opt$f_c\propto t^{-3/2}$ holds in this regime, which will be confirmed in \eqn{series-optfc}.
 Assuming that we will be working in the vicinity of opt$f_c$, it is required to take this scaling into account
 to preserve terms of the right order in the expansion. Therefore, we define the prefactor $p_c$ via 
\eq{pc}{
f_c = \frac{p_c}{t^{3/2}}.
}

\subsection{Chemical potential and related quantities}
To obtain an approximation to $\mu$, the equation of state \eqn{mu-M} is first evaluated to the form
\eq{muint-est}{
\frac{1}{t} = \frac{1}{u_G} + \frac{1}{\pi}\sqrt{\frac{\gamma}{t\,u_G}}\left[1-\frac{\pi^2u_G^2}{12}+\frac{\pi^4u_G^4}{48}-\frac{\pi^6u_G^6}{32}+\mc{O}(u_G^8)\right].
}
In detail, the integral in \eqn{mu-M} can be written as
\eq{integ1}{
-\frac{T}{2\pi\sqrt{\mu}}\int_0^{\infty} ds\left[\frac{s}{R(e^{s R \sqrt{2/u_G}}-1)} + \frac{s-R\sqrt{2/u_G}}{2R}\right]
}
with $R=\sqrt{1+\frac{1}{2}s^2u_G}=\sqrt{1+\Delta(s)}$. While the second term easily integrates, the first does not.
 However, the $[e^{s R\sqrt{2/u_G}}-1]^{-1}$ factor cuts out any contributions at large $s\gtrsim\sqrt{u_G/2}$.
 Since $u_G\ll1$, then $\Delta$ takes on small values $\Delta\lesssim \frac{1}{4}u_G^2\ll1$, and $R\approx1$.
 The first term in the integrand of \eqn{integ1} can be written as a MacLaurin series in $r(s)=R-1\ll1$ as 
$\sum_j M_j(s) r(s)^j$. 
Its integration is still troublesome beyond the lowest terms, 
so $r(s)$ is further expanded in the small quantity $s^2u_G/4$ like 
$r=\frac{s^2u_G}{4}-\frac{s^4u_G^2}{32}+\dots$. This then gives the following series 
\eq{integ2}{
-\frac{T}{2\pi\sqrt{\mu}}\sum_{jj'>0}\int_0^{\infty} ds \mc{T}_{jj'}(s)\ \left(\frac{s^2u_G}{4}\right)^{j'} +\frac{\sqrt{\gamma}}{\pi\sqrt{t\,u_G}},
}
in which all the integrals  give \eqn{muint-est}. We spare the reader from explicit expressions for the $\mc{T}_{jj'}$.

Now, to obtain a self-consistent expansion for $\mu$, we postulate an ansatz
\eq{mu-ansatz}{
\mu = \frac{k_BT}{t}\sum_{j,j'\ge0} c_{jj'}\,t^j\gamma^{j'/2}
}
with coefficients $c_{jj'}$ to be determined. 
By equating subsequent terms of the same orders of $\sqrt{\gamma}$ and $t$ appearing in \eqn{muint-est} one obtains the following series expansion:
\eqa{mu-est}{
\frac{\mu}{k_BT} &=& \frac{1}{t}\left[1-\frac{\sqrt{\gamma}}{\pi}+\frac{\gamma}{2\pi^2}-\frac{\gamma\sqrt{\gamma}}{8\pi^3}\right]\hspace*{7em}\nonu\\
&&+ \frac{t\sqrt{\gamma}\,\pi}{12}\left[1+\frac{\sqrt{\gamma}}{\pi}+\frac{3\gamma}{8\pi^2}\right]
 \mc{O}(t^3,\gamma^2).
}

For c-fields, the integral in \eqn{mu-C} can be expresed as
\eq{estate-C}{
\frac{1}{t} = \frac{1}{u_G^{\rm(cf)}} - \frac{\sqrt{\gamma t u_G^{\rm(cf)}}}{\pi}\, {\rm tan}^{-1}\left[f_c\sqrt{\frac{\pi u_G^{\rm(cf)}}{2}}\right],
}
and the resulting series expansion is
\eqa{mu-est-C}{
\frac{\mu^{\rm(cf)}}{k_BT} &=& \frac{1}{t} + \frac{\sqrt{\gamma}}{2}\left[1-\frac{t\sqrt{8}}{p_c\pi^{3/2}}+\frac{4t^3\sqrt{2}}{3\pi^{5/2}p_c^3}\right]
- \frac{\gamma t}{8}\left[1-\frac{t\sqrt{8}}{\pi^{3/2}p_c}\right]\nonu\\
&&+ \frac{5\gamma^{3/2}t^2}{64}\left[1-\frac{8\sqrt{8}t}{5\pi^{3/2}p_c}\right]
+\mc{O}(t^4,\gamma^2).
}

 The leading correction terms in \eqn{mu-est} and \eqn{mu-est-C}, of $\mc{O}(\sqrt{\gamma})$, have the opposite sign and no cutoff dependence.
 This proves what was previously found empirically:  no cutoff choice will match chemical potentials exactly in the quantum fluctuating regime.
 Since all of $u_G$, $\mc{E}_{\rm int}=T/u_G-T/2t$ and $g^{(2)}(0)=2t\mc{E}_{\rm tot}/T=(2t/u_G)-1$
 depend simply on $\mu$ and the control parameters $\gamma$ and $\tau_d$, they will \emph{never be exactly matched} by any cutoff. 
 The leading order terms for the various observable estimates (which may be of use for future work) are:
\eq{uG-est}{
u_G = t\left[1+\frac{\sqrt{\gamma}}{\pi}+\frac{\gamma}{2\pi^2}\right] -\frac{\pi t^3\sqrt{\gamma}}{12}\left[1+\frac{3\sqrt{\gamma}}{\pi}\right] + \mc{O}(t^5,t\gamma^{3/2})
} 
\eq{uG-est-C}{
u_G^{\rm(cf)} = t -\frac{t^2\sqrt{\gamma}}{2} +t^3\left[\frac{\sqrt{2\gamma}}{\pi^{3/2}p_c}+\frac{3\gamma}{8}\right] + \mc{O}(t^4,\gamma^2)
} 
\eqa{g2-est}{
g^{(2)}(0) &=& 1-\frac{2\sqrt{\gamma}}{\pi}+\frac{\gamma}{\pi^2}-\frac{\gamma^{3/2}}{4\pi^3} + \frac{\pi t^2\sqrt{\gamma}}{6}\left[1+\frac{\sqrt{\gamma}}{\pi}\right]\nonu\\
&&+\mc{O}\left(t^4,t^2\gamma^{3/2},\gamma^2\right)
} 
\eqa{g2-est-C}{
g^{(2)}(0)^{\rm(cf)} &=& 1 + t\sqrt{\gamma} -t^2\sqrt{\gamma}\left[\frac{\sqrt{8}}{\pi^{3/2}p_c}+\frac{\sqrt{\gamma}}{4}\right]\\
&& + t^3\gamma\left[\frac{1}{\sqrt{2}\pi^{3/2}p_c}+\frac{5\sqrt{\gamma}}{32}\right]
+\mc{O}(t^4,\gamma^2).\nonu
}

\subsection{Kinetic energy}
The integral \eqn{ve-M} can be reduced to integrable terms  in the same way as the one in \eqn{mu-M}. Upon substituting \eqn{mu-est} and keeping consistent orders, we obtain
\eqa{eps-est}{
\frac{\ve}{k_BT} &=& \frac{\sqrt{\gamma}}{3\pi t}\left[1-\frac{3\sqrt{\gamma}}{2\pi} + \frac{9\gamma}{8\pi^2} + \frac{\pi^2 t^2}{4}\left( 1 + \frac{\sqrt{\gamma}}{\pi} + \frac{3\gamma}{8\pi^2}\right)\right]\nonu\\
&&+\mc{O}(t^3,\gamma^2).
}
The integral in \eqn{ve-C} can be performed, so
\eq{}{
\frac{\ve^{\rm(cf)}}{k_BT} = 
\sqrt{\frac{2\gamma t}{\pi}}\left(f_c - \frac{1}{\sqrt{2\pi u_G^{\rm(cf)}}}\, {\rm tan}^{-1}\left[f_c\sqrt{\frac{\pi u_G^{\rm(cf)}}{2}}\right]\right).
}
This leads to the following expressions:
\eqa{eps-est-C}{
\frac{\ve^{\rm(cf)}}{k_BT} &=& \frac{p_c\sqrt{2\gamma}}{t\sqrt{\pi}} -\frac{\sqrt{\gamma}}{2} + \frac{t\sqrt{2\gamma}}{\pi^{3/2}p_c} -\frac{t\gamma}{8} + \frac{3t^2\gamma}{p_c\sqrt{8}\pi^{3/2}} \nonu\\
&&+ 
\mc{O}(t^3,t^2\gamma^{3/2},\gamma^2)
}
with the discrepancy 
\eqa{rms-eps}{
\delta_{\ve} &=& 3p_c\sqrt{2\pi}-1 +\frac{9p_c\sqrt{\gamma}}{\sqrt{2\pi}} + \frac{27p_c\gamma}{4\sqrt{2}\pi^{3/2}} 
 - t^2\frac{3(p_c^2\pi^3-4)}{p_c\sqrt{8\pi}} \nonu\\
&&- \frac{3\pi t}{2}\left[1+\frac{3\sqrt{\gamma}}{2\pi}+\frac{9\gamma}{8\pi^2}\right]+ \mc{O}(t^2\sqrt{\gamma},\gamma^{3/2},t^3).
}
Equating this to zero gives the optimum cutoff for kinetic energy (only):
\eqa{eigenfc-eps}{
f_c^{(\ve)} &=& \frac{1}{3\sqrt{2\pi}\,t^{3/2}}\Bigg[1-\frac{3\sqrt{\gamma}}{2\pi}+\frac{9\gamma}{8\pi^2}+\frac{3\pi t}{2} 
+ \mc{O}(t^2,\gamma^{3/2})
\Bigg].
\nonu\\
}
One can see that the leading factor  of \eqn{eigenfc-eps}
 when converted to $\gamma$, $\tau_d$ variables is
$f_c^{(\ve)} = \frac{1}{12\pi^2}(\gamma/\tau_d)^{3/2}(1+\dots)$. The prefactor  $\frac{1}{12\pi^2}= 0.00844$ is a remarkably close match to that seen in \eqn{fc-estmp}. 
To include the other observables and get an estimate for min$RMS$, analysis of the full $RMS$ figure of merit is necessary.

\subsection{Analytic optimization}
\label{S:BOGRMS}
The discrepancy for total energy is 
\eqa{rms-Etot}{
\delta_{\mc{E}_{\rm tot}} &=& 
\sqrt{\gamma}\left[\frac{4}{3\pi}+2p_c\sqrt{\frac{2}{\pi}}-\frac{\pi t^2}{3}\right] 
+ \gamma\left[\frac{16}{9\pi^2}+\frac{8\sqrt{2}p_c}{3\pi^{3/2}}\right]\nonu\\
&&-t^2\gamma\left(\frac{31}{18}+\frac{2p_c\sqrt{2\pi}}{3}\right)
 + \mc{O}(t^3,\gamma^{3/2}).
}
Zeroing out the leading term requires negative $p_c$, but $p_c$ must be positive and was assumed $\mc{O}(1)$, so
 $\delta_{\mc{E}_{\rm tot}}$ is always positive in the vicinity of the optimum cutoff that interests us.
In fact, $\delta_{\mc{E}_{\rm tot}}\approx 2\sqrt{\gamma}/\pi$ at the $f_c^{(\ve)}$ cutoff.

As a corollary to the above, the term  $\mc{M} = {\rm max}\Big[ \delta^{\,2}_{\ve}, \delta_{\mc{E}_{\rm tot}}^{\,2} \Big]$ in \eqn{RMSdef} must take on the flat-bottomed shape seen in Fig.~\ref{sfig:rms-qqc}(c). The ends of the flat-bottomed part will occur when $\delta_{\mc{E}_{\rm tot}}=\pm\delta_{\ve}$, i.e when
\eqa{fcleftright}{
f_c^{\pm} &=& \frac{1}{3\sqrt{2\pi}\,t^{3/2}}\Bigg[1+\frac{(\pm4-3)\sqrt{\gamma}}{2\pi}+\frac{(59\mp32)\gamma}{24\pi^2}+\frac{3\pi t}{2} \nonu\\
&&\hspace*{-2em}\pm t\sqrt{\gamma} +\frac{(4\mp1)\gamma t}{6\pi}
+ \mc{O}(\gamma^{3/2},t^2).\Bigg]
\qquad
}

 When the full $RMS(f_c)$ in the flat bottom region is constructed,
 we have $RMS^2=\delta_{u_G}^2+\delta_{\mc{E}_{\rm tot}}^2 = \gamma(\frac{25}{9\pi^2}+\frac{16\sqrt{2}\,p_c}{3\pi^{3/2}}+\frac{8p_c^2}{\pi}+\mc{O}(t,\sqrt{\gamma}))$.
 The leading order of this always has positive gradient with $f_c$. This implies that  the leftmost edge 
 corresponds to the overall minimum of min$RMS$, i.e. opt$f_c=f_c^-$, 
\eqa{series-optfc}{
{\rm opt}f_c &= \frac{1}{3\sqrt{2\pi}\,t^{3/2}}&\Bigg[1-\frac{7\sqrt{\gamma}}{2\pi}+\frac{91\gamma}{24\pi^2}+\frac{3\pi t}{2} - t\sqrt{\gamma} +\frac{5\gamma t}{6\pi}\nonu\\
&&+ \mc{O}(t^2,\gamma^{3/2})
\Bigg].
}
The global figure of merit at this point is
\eqa{series-minRMS}{
{\rm min}RMS &=& 
\frac{\sqrt{5\gamma}}{\pi}\Bigg[
1 + \frac{\sqrt{\gamma}}{30\pi}+\frac{\pi t}{2} + \frac{\sqrt{\gamma} t}{6} -t^2\left(6+\frac{\pi^2}{12}\right)\nonu\\
&&  + \gamma\frac{1889}{1800\pi}+ \mc{O}(t\gamma,t^2\sqrt{\gamma},\gamma^{3/2},t^2)
\Bigg].
}

\vfill

\end{document}